%% file: main.tex
\pdfoutput=1
\documentclass{article}
\usepackage{spconf,amsmath,amssymb,graphicx}
\usepackage[table]{xcolor}
\usepackage{booktabs}
\usepackage{cite} %
\usepackage{algorithm}
\usepackage{algpseudocode} %
\usepackage[hidelinks]{hyperref}

\newif\ifdraft
\draftfalse
\ifdraft
  \newcommand{\reserved}[1]{{\color{red}[#1]}}           %
  \newcommand{\todo}[1]{{\color{purple}\textbf{TODO:} #1}}
\else
  
  \newcommand{\reserved}[1]{\textbf{??}}
  \newcommand{\todo}[1]{}
\fi

\def\x{{\mathbf x}}
\def\y{{\mathbf y}}
\def\z{{\mathbf z}}

\newcommand{\sg}{\operatorname{sg}}

\title{UNSUPERVISED SPEECH ENHANCEMENT VIA DRIFTING}
\name{Diego Caviedes-Nozal$^{1}$, Liang Xu$^{2}$, Rasmus Kongsgaard Olsson$^{1}$,
      W. Bastiaan Kleijn$^{2}$}
\address{$^{1}$GN A/S, Denmark \quad
         $^{2}$Victoria University of Wellington, New Zealand}

\usepackage{enumitem}

\begin{document}
\ninept
\maketitle

\begin{abstract}
This paper addresses unsupervised speech enhancement in the unpaired setting using drifting methods, where training relies on separate collections of degraded and clean audio without corresponding pairs. While recent drifting approaches enable unpaired training, they do so at a heavy cost: because the objective optimizes only a marginal prior over clean speech, the enhancer gradually loses the input's linguistic content and speaker identity. To fix this, we introduce \emph{input-conditioned drifting}. We preserve the pull of the clean corpus while re-tethering the output to the degraded input via two mechanisms: an \emph{anchor} encoder supplies the missing likelihood by pulling toward the input's features, and a \emph{key} encoder conditions the prior by re-weighting retrieved frames. Neither requires labels or paired data. Using a training-free encoder selection criterion, Word Error Rate on VoiceBank--DEMAND falls to 10.1\% (unprocessed: 11.7\%), speaker similarity recovers from 0.490 to 0.879, and the recipe transfers in part to dereverberation on WSJ0-REVERB: content improves, rendering quality does not.
\end{abstract}


\input{body}

\vfill\pagebreak

\bibliographystyle{IEEEbib}
\bibliography{refs}

\end{document}

%% file: body.tex
\section{Introduction}
\label{sec:intro}

Supervised speech enhancement algorithms \cite{wang2023tfgridnet,richter2023speech} are trained on simulated pairs of degraded audio and clean reference speech \cite{valentini2016vbdmd,reddy2020dns}. In real-world deployment such pairs do not exist: room and device acoustics, ambient noise, and transmission channels alter utterances irreversibly, making the clean version of a real recording unobtainable. Methods that avoid this requirement are called \emph{unsupervised}.

There are several unsupervised approaches in the literature. The first trains on clean speech alone, learning a generative prior that an explicit model of the degradation turns into an enhancer at inference \cite{nortier2024udiffse,ayilo2026diffuseen}. A second trains on degraded audio alone, supervised by a learned quality predictor \cite{fu2022metricganu,jiang2025mosgan} or by re-corrupting the observation into noisier-to-noisy regression pairs \cite{jiang2026selfse}; a related line exploits the statistics of mixtures by remixing them \cite{wisdom2020mixit,tzinis2022remixit}. We address a third regime, \emph{unpaired} training, which uses both collections deployment provides, a system's own degraded inputs and a separate clean corpus, with no correspondence between them \cite{yuan2020cyclegan,yu2021cincgan,jiang2023unse}.

In unsupervised learning, the difficulty is not data volume but correspondence. Collections determine only the marginal distributions of degraded, noise and clean speech, and marginals do not determine which clean signal belongs to a given degraded one. Existing methods therefore add a constraint: a reconstruction requirement \cite{yuan2020cyclegan,yu2021cincgan} or a transport assumption \cite{jiang2023unse}. 

Drifting \cite{deng2025drifting} is a recent generative method that constrains the correspondence in a different way. It iteratively displaces generated samples toward clean examples in a suitable feature space, repeating the process until the generated distribution matches the clean one. Therefore, drifting does not imply any assumption on the corruption mechanism. DriftSE \cite{xu2026driftingmodels,liang2026driftse} applies drifting to speech enhancement showing competitive results when pairs are available. In the unpaired setting, however, it tends to modify linguistic content while maintaining non-intrusive quality.

We investigate this failure mode. Because the standard drift objective is decoupled from the degraded input, and its feature-space gradients may misalign with the needed waveform correction, it loses the input's speaker identity and linguistic content. 

Our contributions are threefold: (1) a Bayesian diagnosis of unpaired content drift; (2) \emph{input-conditioned drifting}, which re-tethers the target to the input using \emph{anchor} and \emph{key} encoders without requiring paired data; and (3) a self-paired, GAN-based corrector that aligns the feature-space update with the waveform. We also provide a training-free criterion for encoder selection in each of the three contributions. Together they make ours the first drifting system that enhances speech without paired data: the Word Error Rate (WER) of the unprocessed input falls from 11.7\% to 10.1\% in a single inference step, identity remains under explicit control, and the recipe transfers in part to dereverberation.

\section{Method}
\label{sec:method}

\subsection{The content-drift problem in unpaired drifting}
\label{ssec:erosion}

Let $\y\in\mathbb{R}^{L}$ be a degraded recording and $\x\in\mathbb{R}^{L}$ its clean counterpart. A generative enhancer takes a prior noise sample $\varepsilon\sim p_\varepsilon$ and the degraded input $\y$, and produces an estimate $\hat\x=f_\theta(\varepsilon,\y)$, where $f_\theta$ is the generator to be trained. The enhancement goal is to match the conditional distribution $p(\x\mid\y)$. Drifting \cite{deng2025drifting} pushes the generated distribution toward the clean one along a Wasserstein gradient flow \cite{han2026wflow}. This flow is computed in a feature space. An \emph{encoder} $\phi:\mathbb{R}^{L}\to\mathbb{R}^{T\times D}$ maps a waveform to a sequence of $T$ feature vectors of dimension $D$, $\z_t=\phi(\hat\x)_t$, referred to as \emph{frames}.

Across a batch of $B$ recordings, the resulting $BT$ frames are treated as an exchangeable set: each generated frame is attracted to clean positive frames $\z^{+}_k$ and repelled by the other frames $\z_{t'}$ in the batch, giving the empirical velocity field
\begin{equation}
  \label{eq:drift}
  V(\z_t)=\sum_{\tau}\Big[\sum_{k} w^{+,\tau}_t(\z^{+}_k)(\z^{+}_k-\z_t)
     -\lambda\sum_{t'} w^{-,\tau}_t(\z_{t'})(\z_{t'}-\z_t)\Big],
\end{equation}
using kernel affinities $w^{\pm,\tau}_t(\z')\propto e^{-d(\z_t,\z')/\tau}$ evaluated at multiple temperatures $\tau$. The generator is then trained to match this displaced target via the loss function:
\begin{equation}
  \label{eq:loss}
  \mathcal{L}_{\text{drift}}=\mathbb{E}\,\big\|\z_t-\sg(\z_t+V(\z_t))\big\|^2 ,
\end{equation}
where $\sg$ is the stop-gradient operator.

\label{ssec:fixedpoint}%
Without paired data, the positive frames cannot come from the utterance's own clean counterpart; they are drawn instead from a generic clean bank $\mathcal{B}=\{b_j\}$, frames of the clean corpus computed by the \emph{bank encoder} $\phi$; we call its entries \emph{rows}. The attraction term in \eqref{eq:drift} therefore pulls each frame toward
$T(\z_t)=\sum_j w^{+}_t(b_j)\,b_j$, the bank's local weighted mean around $\z_t$.

Training halts when the field vanishes ($V\approx0$). The stationarity condition depends only on feature-space locations, so the objective cannot distinguish which degraded input produced a frame: a fluent output attains the same loss whether it carries its own input's words and voice or another's. Moreover, the kernel-smoothed bank density contains no structure below its radius $\tau$, so any attribute that the encoder does not resolve---e.g. fine timbre cues---is unconstrained at equilibrium and averages across the bank. We call this \emph{content drift}: the output remains clean speech, but it no longer preserves the input's linguistic content and speaker identity.

\textbf{A Bayesian Diagnosis.} Framed probabilistically, generative speech enhancement requires sampling from the posterior of a clean frame given its observation:
\begin{equation}
  \label{eq:posterior}
  \log p(\z\mid\y)=\log p(\y\mid\z)+\log p(\z)+\mathrm{const}.
\end{equation}
Standard drifting models only the prior. Under a Gaussian kernel, the attraction target $T(\z_t)$ from \eqref{eq:drift} matches the score (the gradient of the log-density) of a kernel density estimate for the clean marginal prior, $\hat p(\z)\propto\sum_j e^{-d(\z,b_j)/\tau}$:
\begin{equation}
  \label{eq:score}
  \tfrac{\tau}{2}\,\nabla_{\z}\log\hat p(\z_t)=T(\z_t)-\z_t .
\end{equation}
The affinities act as a posterior over \emph{which bank row} produced $\z_t$ \cite{turan2026scorematching}. The repulsion term provides the corresponding score for the generated distribution $q_\theta(\z)$, preventing collapse. The likelihood term $\log p(\y\mid\z)$ is missing: the objective estimates only a marginal prior, and no force ties the output to its own degraded input $\y$.

\label{ssec:jacobian}%
\textbf{Feature-Space Misalignment.} A second structural issue dictates \emph{how} the generator is updated. Differentiating $\mathcal{L}_{\text{drift}}$ with the target held fixed reveals that the gradient reaching the waveform is the velocity field mapped back through the encoder's transposed Jacobian, $J_\phi^{\!\top}$. Even a feature space that separates clean from degraded speech well can push the waveform in a destructive direction. We can score an encoder's suitability by evaluating a known pair $(\x,\y)$. Letting $r=\phi(\x)-\phi(\y)$ be the feature residual and $\delta=\x-\y$ be the ideal waveform correction, the alignment $a(\phi)$ is:
\begin{equation}
  \label{eq:jacobian}
  \nabla_{\hat\x}\mathcal{L}_{\text{drift}}\propto-J_\phi^{\!\top}V(\z_t),
  \qquad a(\phi)=\cos\!\big(J_\phi^{\!\top}r,\,\delta\big).
\end{equation}
Because the bank, temperatures, and batch size act entirely through $V(\z_t)$, no reweighting of the objective can fix a poorly aligned $J_\phi$.

This analysis makes five predictions, which we test in Sec.~\ref{sec:results}: (i) unresolved attributes degrade monotonically while resolved ones improve; (ii) any input-blind objective behaves this way; (iii) reinstating a dependence on the input slows or stops the drift; (iv) an encoder that barely separates clean from degraded speech exerts almost no force; and (v) the distance an input-derived target sits from other utterances predicts whether it transfers.

\subsection{Input-conditioned drifting: anchors and keys}
\label{ssec:onelaw}

\begin{figure}[t]
  \centering
  \includegraphics[width=\columnwidth]{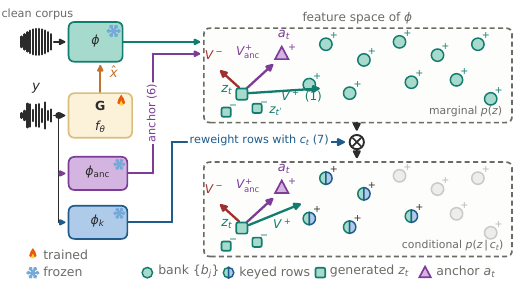}
  \vspace{-1.6em}
  \caption{The unpaired training recipe. The generator $f_\theta$ ($\mathbf{G}$) maps the noisy input $\y$ to $\hat\x$. In the feature space of the bank encoder $\phi$, a generated frame $\z_t$ is attracted by clean bank rows $\{b_j\}$ ($V^+$) and repelled from the batch's other generated frames $\z_{t'}$ ($V^-$). Two input-conditioned mechanisms: the \textbf{anchor encoder} $\phi_{\text{anc}}$ provides a direct target $a_t$ and force $V_{\text{anc}}^+$; the \textbf{key encoder} $\phi_k$ reweights $\{b_j\}$, yielding a conditional prior $p(\z \mid c_t)$ that corrects $V^+$.}
  \label{fig:recipe}
  \vspace{-1.0em}
\end{figure}

To repair the input-blindness of Sec.~\ref{ssec:erosion}, we re-inject the degraded input into the force law (Fig.~\ref{fig:recipe}) without relying on paired data. In the posterior \eqref{eq:posterior} we can re-inject the input at the missing likelihood, or by conditioning the prior. These correspond to the two components \eqref{eq:drift} is built from: the target set and the retrieval weighting.

\textbf{Anchoring to the input.} We supply the missing likelihood by introducing an \emph{anchor}. A secondary encoder $\phi_{\text{anc}}$ maps the signal into content and speaker embedding spaces \cite{chen2022wavlm,desplanques2020ecapa}. We restrict the positive target set for this encoder to just the degraded input frame itself, $a_t=\phi_{\text{anc}}(\y)_t$. The anchor displacement becomes:
\begin{equation}
  \label{eq:anchor}
  V_{\text{anc}}(\z_t)=a_t-\z_t .
\end{equation}
A frame is now exposed to the combined pull of $V(\z_t)+\alpha V_{\text{anc}}(\z_t)$, where $\alpha$ is the anchor weight: the output is pulled directly toward the input's own words and voice. Eq.~\eqref{eq:anchor} is the gradient of a Gaussian likelihood in the anchor's space, a valid observation model only if the corruption barely moves that space ($\phi_{\text{anc}}(\y)\approx\phi_{\text{anc}}(\x)$).

\textbf{Keying (conditioning the prior).} As a key looks up entries in a table, \emph{keying} retrieves the bank rows that resemble the degraded input: a key encoder $\phi_k$ embeds the input as $c_t=\phi_k(\y)_t$, and rows near it are up-weighted ($b^{k}_j$: row $j$ under $\phi_k$):
\begin{equation}
  \label{eq:key}
  w^{\beta}_t(b_j)\propto
  \exp\!\Big(-\frac{d(\z_t,b_j)}{\tau}-\beta\,\frac{d(c_t,b^{k}_j)}{\tau_c}\Big).
\end{equation}
Multiplying the original kernel by a second kernel centered on the input shifts the estimate from the marginal prior to a conditional prior $p(\z\mid c_t)$, with $\beta$ controlling the conditioning strength. For any $\beta$ the target remains a convex combination of clean bank rows, so it is assembled from input-like clean speech: unlike the anchor, which pulls toward the input's own coordinates, the key lets residual noise act only by retrieving wrong rows. %

\label{ssec:seats}%
\textbf{Choosing the three encoders.} The bank, anchor, and key roles may each use a different encoder ($\phi$, $\phi_{\text{anc}}$, $\phi_k$). They place distinct, sometimes opposing, demands: probing frozen representations is standard practice \cite{pasad2023layerwise}, but a single global ranking will not suffice \cite{aldeneh2025rank}. The \emph{bank} must separate clean from degraded speech, or it exerts no pull. The \emph{anchor} must do the opposite, remaining robust to corruption so the degraded input is a safe target. The \emph{key} must retrieve the same rows regardless of degradation. Invariance alone is insufficient: an encoder that maps every voice to a single point is invariant but useless. The criterion is a \emph{margin}: the degraded utterance must be clearly closer to its own clean version than to other utterances. We screen each role before training (Table~\ref{tab:witness}; Sec.~\ref{ssec:res_dereverb}).

\begin{figure}[t]
  \centering
  \includegraphics[width=\columnwidth]{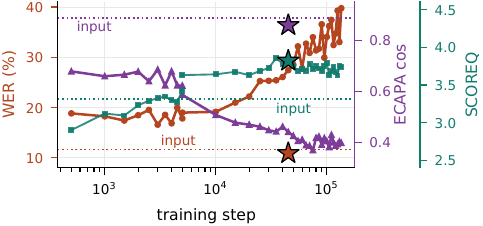}
  \caption{Content and speaker-identity drift of the baseline trained with the input-blind objective
  (VoiceBank--DEMAND test set, WavCube bank). WER and speaker similarity deteriorate throughout training; no checkpoint beats the unprocessed input (dotted). SCOREQ (far right axis) is flat past 5k and sits above its input: quality says nothing about training time. $\bigstar$: the proposed system ($\beta{=}1$ at 45k, Table~\ref{tab:denoise}), one per metric in its colour. Training protocol in Sec.~\ref{ssec:res_erosion}.}
  \label{fig:erosion}
\end{figure}

\begin{table}[t]
  \centering
  \caption{Screening encoders before training (WSJ): phone separability (linear
  probe at PCA-128), recoverability (ridge $R^2$ of the clean log-spectrum), and
  the alignment $a(\phi)$ of \eqref{eq:jacobian} (chance ${\sim}0.06$), against the
  WER and PESQ of training one system per row with that latent as its bank, as
  reported in \cite{liang2026driftse}. Phone separability orders WER,
  recoverability PESQ; alignment neither.}
  \label{tab:witness}
  \small
  \setlength{\tabcolsep}{4pt}
  \resizebox{\columnwidth}{!}{%
  \begin{tabular}{llccccc}
    \toprule
    & & \multicolumn{3}{c}{diagnostics (training-free)} & \multicolumn{2}{c}{outcome} \\
    \cmidrule(lr){3-5}\cmidrule(lr){6-7}
    latent & family & phone-sep & recov.\ $R^2$ & Jac.\ align & WER$\downarrow$ & PESQ$\uparrow$ \\
    \midrule
    PANNs   & tagger & 0.16 & 0.46 & +0.01 & 17.7 & 1.61 \\
    BEATs   & tagger & 0.36 & 0.70 & +0.06 & 9.9 & 1.84 \\
    DistilHuBERT & SSL & 0.60 & 0.86 & \textbf{+0.11} & 5.5 & 2.03 \\
    WavLM-large & SSL & \textbf{0.68} & \textbf{0.87} & +0.06 & 5.3 & 2.18 \\
    WavCube & vocoder & 0.65 & 0.83 & \textbf{+0.11} & 5.0 & 2.30 \\
    \bottomrule
  \end{tabular}}
\end{table}

\section{Experiments and results}
\label{sec:results}

After the setup, we test the analysis of Sec.~\ref{sec:method}: content drift over training, the two improvements and the key's cost, a single-pipeline comparison, transfer to a second corruption, and the corrector. Prediction (ii), which our ablations cannot isolate, is addressed in Sec.~\ref{sec:discussion}.

\subsection{Setup}
\label{ssec:setup}
\textbf{Data.} Denoising uses VoiceBank--DEMAND \cite{valentini2016vbdmd}
(10{,}802 train, $770$ validation, complete $824$-utterance test set) and
dereverberation WSJ0-REVERB \cite{garofolo1993wsj0} (rooms with $T_{60}$ uniform in $0.4$--$1.0$\,s, anechoic targets from the same
geometry \cite{richter2023speech}, $651$-utterance test set), at $16$\,kHz. %
The clean bank is built from DNS-2020 \cite{reddy2020dns} and is disjoint from both evaluation
corpora, so no term of the loss ever sees the clean signal of the utterance being processed.

\textbf{System.} Both tasks share a 1.69M-parameter TF-GridNet \cite{wang2023tfgridnet} generator trained for 45k steps with an effective batch size of 32 (batch 8 with 4-step gradient accumulation). The prior is $\varepsilon\sim\mathcal{N}(0,\sigma^{2}I)$, where $\sigma$ follows a truncated log-normal distribution. Affinities are softmax-normalized along the frame and bank axes using temperatures $\tau\in\{0.02,0.05,0.1\}$ (0.3 for the identity anchor). We use a DNS-2020 bank of 10,240 frames per step. The anchors rely on WavLM-base-plus L6 ($\alpha{=}0.3$) \cite{chen2022wavlm} and ECAPA-TDNN ($\alpha{=}0.05$) \cite{desplanques2020ecapa}. A repository provides code, hyperparameters and audio examples\footnote{\url{https://github.com/d-caviedes/unpaired-drifting-se}; audio: \url{https://d-caviedes.github.io/unpaired-drifting-se}}.

\textbf{Metrics.} We rescore all baselines within our own pipeline (quality: DNSMOS \cite{reddy2022dnsmos}, SCOREQ \cite{ragano2024scoreq}) using the authors' released checkpoints or audio. This is necessary because non-intrusive metrics differ across implementations by over 0.5 DNSMOS on the same unprocessed test set.

\subsection{Content drift over training}
\label{ssec:res_erosion}

Fig.~\ref{fig:erosion} tracks the unanchored recipe (WavCube encoder) over training. Between 5k and 135k steps, WER rises by 13.4$\pm$1.2 points and speaker similarity falls by 0.09$\pm$0.02, both per 100k steps (linear fits), while SCOREQ remains stable. Insertions grow faster than substitutions (6.6$\times$ against 1.8$\times$). SCOREQ instead saturates within the first 5k steps. WER is at its best in that same window, 17--20\%, still worse than the unprocessed input, and only degrades from there.

\begin{table*}[t]
  \centering
  \caption{\textbf{Denoising.} VoiceBank--DEMAND, complete 824-utterance test set. We rescore all methods in our pipeline from one render (external rows: authors' released artefacts). D/I are deletions/insertions; DNSMOS is OVRL; SIG/BAK are quoted in Sec.~\ref{ssec:res_key}. ``Steps'': sampler inference steps (a corrector is a separate net); ``params'': inference-time parameters. Bold: best within each block.}
  \label{tab:denoise}
  \footnotesize
  \setlength{\tabcolsep}{4pt}
  \begin{tabular}{lllcccccccc}
    \toprule
    system & trains on & mech. & steps & params (M) & WER$\downarrow$ (D/I) & ECAPA$\uparrow$ & DNSMOS$\uparrow$ & SCOREQ$\uparrow$ & PESQ$\uparrow$ & STOI$\uparrow$ \\
    \midrule
    noisy input & --- & --- & --- & --- & 11.7 & 0.888 & 2.70 & 3.31 & 1.97 & 0.921 \\
    \midrule
    \multicolumn{11}{l}{\textit{Masking-based}}\\
    \;MetricGAN-U \cite{fu2022metricganu} & mixtures & mask & 1 & 1.9 & 16.1 (116/58) & 0.667 & 2.81 & 3.37 & 2.13 & 0.889 \\
    \;SelfSE \cite{jiang2026selfse} & mix+noise & mask & 1 & 1.5 & \textbf{9.3} (74/37) & \textbf{0.860} & \textbf{3.18} & \textbf{4.13} & \textbf{2.98} & \textbf{0.943} \\
    \midrule
    \multicolumn{11}{l}{\textit{Resynthesis-based}}\\
    \;DiffUSEEN \cite{ayilo2026diffuseen} & clean+noise & diff. & 30 & 5.2 & 10.7 (83/37) & 0.865 & 3.05 & 3.79 & \textbf{2.67} & \textbf{0.933} \\
    \;DriftSE unpaired \cite{liang2026driftse} & clean+mix & drift & 1 & 4.3 & 22.5 (152/285) & 0.569 & 3.10 & 3.73 & 1.73 & 0.609 \\
    \;ours: no anchors (20k) & clean+mix & drift & 1 & 1.7 & 19.6 (131/95) & 0.490 & 2.91 & 3.90 & 1.71 & 0.616 \\
    \;\;+ content anchor & clean+mix & drift & 1 & 1.7 & 14.6 (89/71) & 0.539 & 2.95 & 3.95 & 1.84 & 0.606 \\
    \;\;+ identity anchor (45k) & clean+mix & drift & 1 & 1.7 & 14.0 (75/83) & 0.748 & 2.91 & 3.87 & 1.96 & 0.613 \\
    \;\;+ self-paired corrector (F7) & clean+mix & drift & 1 & 2.5 & 13.6 (101/82) & 0.739 & 3.09 & 3.97 & 2.04 & 0.660 \\
    \;ours: anchors + key, $\beta{=}1$ & clean+mix & drift & 1 & 1.7 & 10.9 (67/52) & 0.859 & 2.91 & 3.82 & 2.43 & 0.925 \\
    \;ours: anchors + key, $\beta{=}4$ & clean+mix & drift & 1 & 1.7 & \textbf{10.1} (69/49) & \textbf{0.879} & 2.76 & 3.59 & 2.23 & 0.926 \\
    \;ours: anchors + gated key & clean+mix & drift & 1 & 1.7 & 10.5 (73/42) & 0.865 & 2.90 & 3.78 & 2.42 & 0.926 \\
    \;ours: keyed $\beta{=}1$ + corrector & clean+mix & drift & 1 & 2.5 & 11.9 (83/40) & 0.842 & \textbf{3.14} & \textbf{4.03} & 2.47 & 0.919 \\
    \bottomrule
  \end{tabular}
\end{table*}

\subsection{Effect of anchoring and keying}
\label{ssec:res_repair}

Each anchor acts on its own attribute (Table~\ref{tab:denoise}): the content anchor carries the WER reduction, and the identity anchor carries the speaker-similarity recovery. Training behaviour changes with them: the anchored system degrades by 1.8 points from 45k to 135k, against 12.3 for the unanchored one. The anchors slow the content drift more than sixfold but do not stop it. Further adding the key takes WER below that of the unprocessed input with the level of pairing unchanged: the bank is disjoint from the evaluation material and the keys come from the degraded signal alone.

\subsection{Effect of key strength}
\label{ssec:res_key}

Sweeping $\beta$ from 1 to 8 gives a U-shaped WER with its minimum at $\beta{=}4$ (Table~\ref{tab:denoise} reports $\beta{=}1$ and $\beta{=}4$). Word insertions fall from 193 to 117 across the sweep, while repairs decay slowly.

The cost is confined to the background: DNSMOS speech quality (SIG) stays at 3.65--3.81 while background quality (BAK) falls from 3.77 to 2.99. The key computes its affinities on the noisy input, and $\sim$60\% of the rows a frame retrieves change between the clean and degraded renderings of the same utterance, so WER and background trade off between the fixed-$\beta$ endpoints. A per-frame $\beta_t$ avoids this trade-off. During training a frozen CTC recognizer \cite{baevski2020wav2vec2} decodes the noisy input (it is absent at inference), yielding a per-frame confidence $q_t$; on letter frames the gate sets $\beta_t = 1 + 3\,\sigma\big((q_t - 0.89)/0.025\big)$ with $\sigma$ the logistic function, interpolating between the endpoints $\beta{=}1$ and $\beta{=}4$, and on CTC blanks $\beta_t{=}1$. The constants are derived from training-side statistics of the recognizer's confidence, not fitted to the evaluation metric. The confidence separates input-correct from input-wrong words (AUC 0.90--0.92), and the gate reaches 10.51\% WER at the background quality of $\beta{=}1$.

\subsection{Comparison with other unsupervised systems}
\label{ssec:res_external}

Table~\ref{tab:denoise} compares our systems with the unsupervised methods whose trained systems or outputs are publicly available. The masking refiner SelfSE attains the best WER and quality in the table. Among the resynthesis systems, only two improve on the input's WER: DiffUSEEN and the keyed system. Exchanging the bank encoder alone moves our system between the regimes: a phone-selective latent stays close to the input, acoustic latents resynthesise. As a control on prediction (iv) of Sec.~\ref{ssec:jacobian}, we replace the pretrained bank encoder with whitened complex-STFT frames: with no pretrained network shaping the space, the system performs worse than the unprocessed input in WER, speaker similarity and DNSMOS.

\begin{table}[t]
  \centering
    \caption{\textbf{Dereverberation.} WSJ0-REVERB, complete 651-utterance test set, same pipeline
  and judge as Table~\ref{tab:denoise} (wav2vec2-base-960h); WER in percent. The full
  system is two banked kernels plus two conditioning anchors (Sec.~\ref{ssec:setup});
  the ablation removes only the anchors.}
  \label{tab:dereverb}
  \small
  \setlength{\tabcolsep}{3.5pt}
  \resizebox{\columnwidth}{!}{%
  \begin{tabular}{lccccc}
    \toprule
    system & WER$\downarrow$ & ESTOI$\uparrow$ & SRMR$\uparrow$ & SCOREQ$\uparrow$ & ECAPA$\uparrow$ \\
    \midrule
    reverberant input (unprocessed) & 32.71 & 0.452 & 3.45 & 2.74 & \textbf{0.814} \\
    ours: 2 kernels $+$ 2 anchors & \textbf{23.81} & \textbf{0.586} & \textbf{12.93} & 2.37 & 0.745 \\
    \;no anchors & 81.69 & 0.433 & 4.16 & \textbf{3.06} & 0.215 \\
    \bottomrule
  \end{tabular}}
\end{table}

\subsection{Dereverberation}
\label{ssec:res_dereverb}

The margin of Sec.~\ref{ssec:seats} (how much closer a degraded utterance is to its own clean version than to other utterances) predicts before training which anchors remain valid here: on matched pairs ($200$ per corpus) the content-anchor margin falls from 0.205 under noise to 0.102 under reverberation while the identity margin rises from 0.585 to 0.633. The identity anchor should remain a safe target, the content anchor should not; the trained results follow (Table~\ref{tab:dereverb}). Removing the two anchors raises WER by 58 points and reduces speaker similarity by 0.53: the content kernel cannot replace them, and without them the two kernels drift even faster than one. The full system improves on the input in WER, ESTOI and SRMR, but not in SCOREQ. The keying mechanism behaves differently here. Raising its strength ($\beta$ $0\to0.3\to1$) moves every measure monotonically (WER 23.81/25.01/26.00, SRMR 12.93/9.82/7.18, ECAPA 0.745/0.769/0.794): here the key is not the selector it is on denoising; it only keeps the output close to the input, trading dereverberation strength for fidelity. A plausible explanation is where the corruption lies relative to the clean speech distribution: additive noise lies outside it and the key can retrieve against it; reverberation is built of speech and may lie inside it, leaving nothing to select against.
\subsection{Self-paired corrector}
\label{ssec:res_second}
\label{ssec:corrector}%

A converged generator modifies the signal in a fixed way that is largely independent of its input, so matched pairs $(G(\x),\x)$ can be generated from the clean corpus alone. A small GAN corrector is trained on them \cite{kaneko2017gan}: a paired regression preserves the content while an adversarial loss on its output removes the artefact texture. The assumed input-independence is checkable in advance: on denoising, a modulation-spectrum ripple statistic of the artefact's 4--8\,kHz band is 0.663 on both $G(\text{clean})$ and $G(\text{noisy})$. The corrector raises SCOREQ while preserving WER on the anchored generator, and at a cost of 1.0 WER point on the keyed one (Table~\ref{tab:denoise}). The construction fails on dereverberation: that generator leaves clean input almost unchanged, so the pairs contain no artefact to learn from.

\section{Discussion and conclusions}
\label{sec:discussion}

\textbf{What input conditioning provides.} Ours is the only unsupervised system in Table~\ref{tab:denoise} that resynthesises speech and improves on its input, in one step and with 1.7M parameters, compared with 5.2M and 30 steps for the nearest generative competitor. Speaker identity stays close to the input (ECAPA 0.879 against 0.888), and the anchors and key act only in training, so inference is one forward pass.

\textbf{What it does not provide.} Masking systems keep the lead on WER and intrusive quality (SelfSE 9.3\%, Sec.~\ref{ssec:res_external}). On dereverberation the gain is confined to content: WER, ESTOI and SRMR improve on the input while SCOREQ and speaker similarity do not. The corrector raises quality on the anchored denoiser, and costs 1.0 WER point on the keyed one.

\textbf{The analysis generalizes beyond drifting.} As predicted in Sec.~\ref{sec:method}, content and identity erode while quality saturates (Fig.~\ref{fig:erosion}), and conditioning on the input stops the drift (Table~\ref{tab:denoise}). The same prediction explains the reported behaviour of MOS-GAN \cite{jiang2025mosgan}: stripped of its input-dependent loss terms, its intrusive measures (e.g.\ PESQ) degrade while its non-intrusive target (a learned quality predictor) improves: content drift, not a training instability. In our system, a room-acoustics bank encoder reproduces the collapse (42.6\% WER): encoder choice alone can cause it. The effect differs from mode collapse: the marginal can match the clean corpus while each output's link to its own input decays \cite{han2026wflow}.

\textbf{Conclusion.} Unsupervised distribution matching via drifting loses content when its training signal is a feature-space field, not a correspondence with the input. Anchoring the target and keying the retrieval restore that correspondence without pairs, allowing single-step generative enhancement to reduce WER below the input while improving quality. The dereverberation quality gap remains open.